# Chemistry in Protoplanetary Disks

## Address

Dr. Dmitry Semenov

Max Planck Institute of Astronomy

Koenigstuhl 17

D-69117 Heidelberg, Germany

semenov@mpia.de

## Synonyms

circumstellar disks, young stellar objects, ...

## Keywords

circumstellar disks, molecules, chemical reactions, molecular lines, (sub-)millimeter observations, isotopic fractionation, Solar Nebula, radio-interferometers, ionization, accretion, high-energy radiation, ...

## Definition

Protoplanetary disks (PPDs) surrounding young stars are short-lived (~0.3-10 Myr), compact (~10-1000 AU) rotating reservoirs of gas and dust. PPDs are believed to be birthplaces of planetary systems, where tiny grains are assembled into pebbles, then rocks, planetesimals, and eventually planets, asteroids, and comets. Strong variations of physical conditions (temperature, density, ionization rate, UV/X-rays intensities) make a variety of chemical processes active in disks, producing simple molecules in the gas phase and complex polyatomic (organic) species on the surfaces of dust particles.

## Overview

One of the ultimate questions the mankind seeks to understand is the origin of life and its prevalence in the Universe. The crucial part of this complicated topic is to understand at what conditions prebiotic molecules and their simpler "building-blocks" species could be produced and survive in cosmos, in particular, at the verge of planet formation. By virtue of rapid technological advances more powerful observational facilities and robotic missions become available, which allow us to remotely study chemical composition in nearby disks and to collect pristine materials of the Solar system *in situ*, while increasing computer power makes sophisticated chemo-physical disk simulations feasible.

The accretion disks around young Sun-like stars and brown dwarfs possess sufficient material to assemble planetary systems [1, 2]. According to the modern paradigm of the star formation, a collapse of a cold quiescent molecular cloud of several solar masses leads to formation of a dense rotating flattened configuration around a protostar due to angular momentum redistribution, which is enshrouded by a natal envelope and a bipolar outflow. During a few Myr of dynamical evolution much of pre-natal material accretes onto the star, ceasing outflow and making the newly born star visible. Photospheric activity of the star drives intense, variable ultraviolet (UV) and X-ray radiation, leading to steady photo-evaporation and, along with ongoing accretion, to dispersal of remaining disk matter. In dense disk interior conditions are favorable for rapid agglomeration of sub-micron-sized dust grains into larger cm/m-sized pebbles and rocks that sediment toward the disk equatorial plane ("midplane"), eventually forming planets.

In disks, molecules are crucial heating and cooling agents of the gas, while dust grains determine opacity of the medium to the stellar radiation, shaping disk thermal and density structure. Ionization chemistry induced by high-energy radiation and cosmic ray particles (CRPs) determines the strength of the coupling between the gas and magnetic fields, thus controlling disk dynamics globally. Consequently, a strong, time-dependent variation of temperature, density, and dissociating radiation intensity across a PPD makes it a harbor for rich chemistry that proceeds both in the gas phase and on the surfaces of dust grains. In this entry, we concisely outline the major modern observational methods and theoretical paradigms used to investigate disk chemical composition and evolution, and present their most important results. Future directions of research that will be opened with the advent of the Atacama Large Millimeter Array (ALMA) and other forthcoming observational facilities will also be discussed.

# Basic Methodology

Table 1: Diagnostic molecules to study disk physics and chemistry

| Species | Probe | Midplane | Molecular layer | Atmosphere | Inner zone |
|---|---|---|---|---|---|
| $^{12}CO$, $^{13}CO$, $C^{18}O$ | Temperature, surface density | mm | mm | mm | IR |
| $H_2$ | Temperature | - | - | - | IR |
| $NH_3$ | --- | cm | cm | - | - |
| CS, $H_2CO$ | Density | - | mm | - | IR |
| $C_2H$, HCN, CN, OH, $H_2O$, $C_2H_2$ | Photoprocesses | - | mm | - | IR |
| $HCO^+$ | Ionization | mm | mm | - | - |
| $N_2H^+$, $H_2D^+$ | --- | mm | - | - | - |
| $C^+$ | --- | - | - | IR | IR |
| metal ions (e.g., NeII) | --- | - | - | - | IR |
| complex organics (e.g, $H_2CO$) | Surface chemistry | cm, mm | cm, mm | - | IR, mm |
| $DCO^+$, DCN, $H_2D^+$ | Deuterium fractionation | mm | mm | - | - |

Nowadays astronomers have discovered about 150 species (230 including isotopomers) in space [3]. These vary from simple radicals (e.g., CH) to complex molecules like cyanopolyynes (e.g., $HC_{11}N$), polyaromatic hydrocarbons (PAHs; up to $C_{60}$ and $C_{70}$), organic molecules (e.g., $CH_3OH$, HNCO, $HCOOCH_3$) as well as positive (e.g., $HCO^+$) and negative (e.g., $C_8H^-$) ions. Among these species only a handful of molecules have been identified in disks through the IR and (sub-)millimeter spectroscopy: $H_2$, NeII, FeI, OH, $H_2O$, CO (and isotopologues), CN, HCN, DCN, HNC, $C_2H$, $C_2H_2$, CS, $HCO^+$, $H_2CO$, $DCO^+$, $NH_3$ and $N_2H^+$ [4-7]. A multitude of solid materials including amorphous and crystalline silicates, various ices, and organic compounds have also been discovered, partly as meteoritic constituents only.

## (Sub-)millimeter observations

Dusty PPDs are transparent only when observed at wavelengths longward of ~100μm. Since the dominant molecular component of the gas, $H_2$, is not directly observable apart from high-temperature regions around the star [8], other trace species and dust are used to study physical conditions in disks (Table 1). At these wavelengths there are no prominent solid-state features, so the observed continuum emission allows us to estimate dust disk mass and the presence of large, mm-sized grains. Since dust opacities at these wavelengths and dust/gas mass ratio are poorly known, the constrained disk mass are subjected to large uncertainties.

In contrast, the (sub-)millimeter spectral window is rich in molecular rotational emission lines. It is not fully accessible from the ground due to atmospheric absorption, primarily by water and hydroxyl. Single-dish telescopes like the IRAM 30-m antenna (ESO, Spain), the APEX 12-m antenna (ESO, Chile), and the 15-m James Clerk Maxwell Telescope (UK, Hawaii) are used to survey PPDs in lines of potentially detectable species. Due to the lack of spatial resolution, such data allow to constrain disk-averaged amounts of emitting molecules (column densities) and roughly estimate disk orientation.

Sub-arc-second imaging required to resolve PPDs is provided by antenna arrays such as the Plateau de Bure interferometer (France), Submillimeter Array (USA), Combined Array for Research in Millimeter-wave Astronomy (USA) and Nobeyama Millimeter Array (Japan). Due to their limited sensitivity such studies are presently scarce and restricted to outer disk regions (r>30AU). The well-studied systems so far include DM Tau, LkCa 15, AB Aur, and TW Hya. In PPDs most of molecular lines are optically thin and thus their apparent spectra is a complicated combination of molecular distribution, excitation conditions, and radiative transfer through the disk. The molecular spectra possess unique information about disk kinematics and geometry, temperature and density structure, and intensities of dissociating and ionizing radiation.

The disk size and orientation can be constrained directly by the geometry of mapped dust emission and integrated line intensity maps (e. g., $^{13}$CO 1-0), see Fig 1. Next, the PPD kinematics can be recovered by analyzing velocity line maps and position-velocity diagrams of various cuts through the disk. For molecules with high dipole moment and regular level structure (e. g., CS, $H_2$CO, $HCO^+$) multi-transition observations allow constraining density structure. Thermal disk structure is typically derived from the optically thick CO lines tracing different disk heights (e. g., $^{12}$CO 1-0, $^{12}$CO 2-1, etc.). The line profiles of CO isotopologues carry information about both temperature and molecular density distribution, which can be used to derive disk surface density profile. Ratios of line intensities from a saturated molecule and its photodissociation radical (like HCN and CN) allow constraining strength of disk irradiation by high-energy UV photons. However, the larger the molecule, the larger the partitioning of the energy levels and the lower the line intensities, making it harder to detect and resolve its spectra. Also, molecules with complex level structure, like water, have particularly hard spectra to analyze and model. Therefore, to extract full information from interferometric spectra of a generic molecule one has to know or constrain consistently density and thermal disk structure, kinematics, geometry, and molecular abundance distributions, and employ a line radiation transport model to fit the data iteratively. This naturally requires feasible disk physico-chemical models, accurate collisional rates and spectroscopic data, and multi-line, multi-molecule high-quality spectral observations.

## Infrared observations

At IR wavelengths emerge various solid-state features of dust constituents as well as rotational-vibrational and vibrational bands of gas-phase species (due to stretching and bending of molecular bonds). Atmospheric transparency windows permits limited observations from the ground, and space-borne He-cooled telescopes get advantage here. The solid-state bands are either in absorption for inclined disks and cold regions (mostly icy features) or in emission for pole-on disks and warm inner zones (e. g., silicate bands). From analysis of these data we can constrain local temperature and amount of absorbing/emitting material on the line of sight. The ratio between near IR and far IR dust emission is used to indicate an evolutionary phase of a PPD because in an old system dust grains may grow so big in the dense inner region that the near-IR flux disappears, while the far-IR excess emission is still characteristic of cold, sub-micron-sized grains.

The inner, ~1-20 AU planet-forming disk regions have became accessible only recently, with the advent of near- and mid-IR ground-based interferometers (such as the Palomar Testbed and Keck Interferometers) and the launch of space observatories (the Infrared Space Observatory, Spitzer, and Herschel). The IR interferometers do not retain the phase information of the detected signal, and thus require sophisticated analysis of the acquired data based on template disk models, constraining geometry and mass of the hot dust region. The (ro-)vibrational emission/absorption molecular lines and solid-state emission bands of ices as well as a variety of silicates and polyaromatic hydrocarbons (PAHs) have been detected in disks [9-12]. Identified molecules include $H_2O$, $C_2H_2$, HCN, OH, NeII and some other; from their spectra local temperature, FUV intensity and molecular concentrations can be derived.

The major components of dust grains in disks are silicates (e. g., $MgSiO_3$ and $Mg_2SiO_4$), troilite (FeS), metals and oxides, quartz, C-containing compounds, and water/CO ices. The prominent solid-state features at 10 and 18μm arise from Mg-rich amorphous silicates (olivine, pyroxene), while crystalline silicates (forsterite, enstatite) show a multitude of distinct peaks between 10 and 30μm. Their relative strengths and line shapes permit to determine a fraction of high-temperature-processed crystalline silicates to the pristine amorphous silicates, and discern average size of emitting particles. Troilite has a broad feature at ~23μm, water ice has a peak at 3μm, while polar and apolar CO ices have a strong feature at 4.67μm. For minor ices, the shift of the band peak and its shape depends on the presence and the lattice structure of the major mantle ices. Metals do not have any distinct spectroscopic signatures, while only certain carbon-bearing compounds show generic aliphatic and aromatic features, making it difficult to identify their exact composition (which is likely a mixture of hydrogenated amorphous carbons, diamonds, PAHs, and polymerized organic materials). Their presence is revealed by chemical and petrological analysis of meteorites and collected interplanetary dust particles and by mass-spectroscopy of cometary grain constituents in our Solar system.

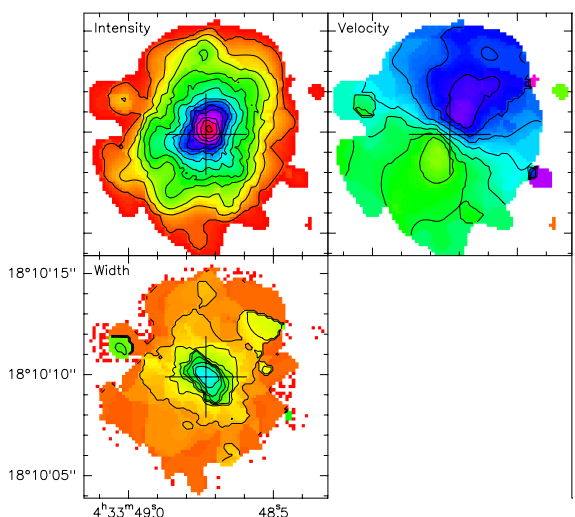

Figure 1. The DM Tau disk imaged with the Plateau de Bure interferometer in the HCO$^+$ 1-0 line with the 1.5" resolution. Shown are the integrated line intensity (top left) that increases toward the central star (indicated by cross), the velocity map (right top) that reveals a "butterfly"-like symmetric pattern typical of rotating disks, and the line width distribution (bottom left), where the line width increases toward the disk center due to Keplerian rotation ($V(r) \propto r^{-1/2}$).

## Primitive materials in the Solar system

The evolutionary history of the Solar Nebula during the first few Myr has been imprinted into compositional variety of meteoritic and cometary materials [13-15]. (Micro-)meteorites routinely retrieved in Antarctica, and cometary grains collected in the Earth atmosphere and in the Solar System by spacecrafts like *Stardust*, as well as bulky meteorites (such as *Murchison* and *Orguel*) are used as samples of these pristine (and partly altered) constituents. Their bulk chemical composition is analyzed by various techniques such as wet chemistry where meteoritic matter is dissolved in various acids to separate the elements based on their chemical properties, X-ray fluorescence analysis of powdered samples, and neutron activation analysis of samples in a nuclear reactor to produce short-lived radioactive species, emitting characteristic gamma rays. By comparing abundances of volatile and heavy elements in Solar photospheres and in meteorites, chondrites has been identified as containing the most primitive solids condensed out in the Solar nebula. Micro-manipulation techniques are being developed to ease the analysis of particularly fragile carbon-rich samples. The (Raman) IR gas-phase spectroscopy and mass-spectroscopy of heated samples is applied to identify individual chemical species.

The isotopic diversity of various types of meteorites is indicative of different ages of their condensation and previous irradiation and heating history, pointing to partly non-homogeneous composition in the inner Solar nebula. By exotic isotopic abundances pre-solar grains, which have formed prior and survived the formation of Solar system, have also been discovered as individual constituents of meteoritic matter. Their characterization is done by non-destructive methods such as scanning electron microscope (SEM) applied to discern their morphological properties and second ion mass spectrometer (SIMS) technique, where high-energy ions are utilized to ionize atoms from the sample to enable their detailed mass spectroscopic studies.

The petrological analysis of carbonaceous meteorites is used to understand the degree to which they have been thermally processed and altered.

In contrast, mineralogical and chemical composition and properties of pristine ices and volatile materials that have not survived during their journey to Earth can only be studied by spectroscopic observations and analysis of gasses in vicinity of cometary nuclei, either by ground-based radio-interferometry when a comet is approaching the Sun or by a fly-by of a satellite probe through cometary coma or by making an artificial impact with the cometary surface leading to a short-time burst of outgassing. For the first time the *Stardust* mission has returned to Earth particles collected directly from Wild 2 comet by aerogel collectors. Some of indigenous organic compounds have been recovered in Martian meteorites and found in atmospheres of Titan and Jovian planets. Less volatile, likely thermally processed organics having various molecular weight, structure, and composition is found in chondrites. These include aliphatic and aromatic (PAHs) hydrocarbons, carboxylic acids, amino acids, amines, alcohols, aldehydes, ketones, and other species [16].

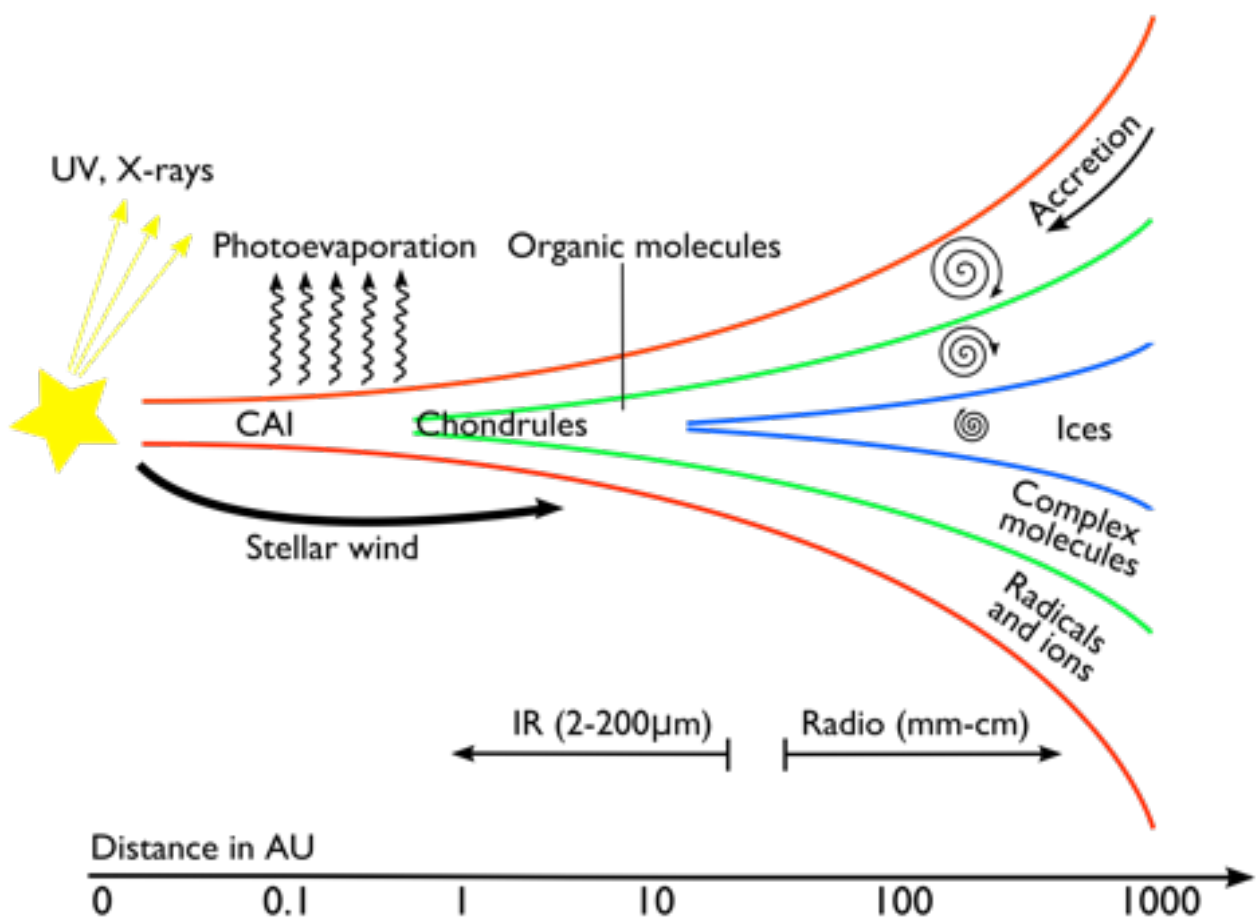

Figure 2. A sketch of the vertical structure of a protoplanetary disk.

# Key Research Findings

### Outer disk: (sub-)millimeter observations
By (sub-)millimeter high-resolution interferometric observations the chemical composition and physics of outer regions (beyond 30-50 AU) in several nearby PPDs have been examined. Observations at the Very Large Array (USA) and the Australia Telescope Compact Array at millimeter and centimeter wavelengths show no evidence for rapid decline of dust continuum intensity with wavelength, indicative of significant grain growth up to cm sizes [17-18]. The IR spectroscopic surveys of young stars in stellar clusters of various ages have placed tight constraints on typical dispersal time of dense protoplanetary disks, ~5 Myr, while their inner, planet- forming zones (<10-100 AU) are cleared of sub-micron-sized dust grains within ~1 Myr only.

The prominent low- and high-lying CO rotational lines are readily excited at densities of $\sim 10^3$-$10^5$ cm$^{-3}$, despite a low dipole moment of CO (0.112 Debye). The $^{12}$CO lines are optically thick and their intensities measure kinetic temperature in the disk upper layer. The lines of less abundant $^{13}$CO and C$^{18}$O are typically optically thin or partially optically thick and are sensitive to both temperature and CO column densities through the disk. Strong CO lines are suitable for accurate determination of disk kinematics, orientation and geometry, and the mass of the central star. From interferometric observations it has been found that measured disks radii are the smallest for the dust continuum and progressively larger in the C$^{18}$O, $^{13}$CO, and $^{12}$CO lines, with a typical value of $\sim$100-1000 AU. Multi-line CO studies revealed a clear sign of vertical temperature gradient in several disks, increasing from $\sim$10 K at the midplane to $\sim$50 K in the atmosphere region [7, 19, 20]. The high surface temperature of the TW Hya disk as measured by the $^{12}$CO (6-5) line cannot be solely caused by the black body stellar radiation and requires an additional heating source, likely stellar X-rays [21]. A number of protoplanetary disks with large inner "dust" holes, such as LkCa15, do not show temperature variation in the vertical direction, and thus shall have peculiar structure. A large amount of CO and other gas-phase molecules at $\sim$10-15 K have been detected in DM Tau disk, which contradicts to the fact that most of these molecules shall reside on grains at such low temperatures.

Another easily observable molecular species in disks is HCO$^+$ (dipole moment is 3.92 D). The low-lying 1-0, 2-1, and 3-2 transitions of this ion are excited at densities of $\gtrsim 10^5$ cm$^{-3}$ and are (partially) optically thin. This is one of the most abundant charged species in PPDs, the other being C$^+$ that is not observable at millimeter wavelengths. Due to large dipole moment, HCO$^+$ is a good density probe, in addition to being a tracer of ionization degree (see Table 1). Another, less abundant observable ion in disks is N$_2$H$^+$ (3.37 D). The 1-0 N$_2$H$^+$ line is hyperfine-splitted, so from relative intensities of its individual components one can reliably determine optical depth of the line and constrain N$_2$H$^+$ column density. Using these ions as probes, it has been found that the degree of ionization in disk interiors is $\sim 10^{-9}$ (one charged species per billion), in agreement with derived cosmic ray ionization rate and predictions from disk chemical models [22].

The distribution and total amount of HCN (2.98 D) and its dissociation product, CN (1.45 D), depends on the intensity and spectral shape of the high-energy UV radiation impinging into the disk [23]. The stronger the FUV flux, the higher the CN to HCN line ratio due to higher concentrations of CN and lower abundances of HCN. The observationally inferred elevated ratio of CN to HCN abundances indicates that chemistry in PPDs is indeed affected by the stellar FUV radiation. Another radical observed in disks, C$_2$H (0.8 D), is sensitive to the X-ray luminosity of the central star, as higher flux of X-ray photons replenishes more of elemental carbon locked in CO back into the gas-phase, where it is quickly converted to light hydrocarbons [24].

Rotational lines of less abundant CS (1.98 D) and H$_2$CO (2.3 D) are difficult to observe even in bright disks. The CS lines, excited at n $\sim 10^5$-$10^7$ cm$^{-3}$, have narrow intrinsic widths and can be utilized to measure turbulent gas velocities in disks, as well as to discern its density structure. Measured turbulent broadening is subsonic, with a typical microturbulent velocity of $\sim$0.05-0.2 km s$^{-1}$. The simplest organic molecule, H$_2$CO, has a slightly asymmetric top structure and serves as both densitometer and thermometer. It is partly produced on surfaces of dust grains and can also be used as a probe of efficiency of surface chemical processes. Recently detected DCO$^+$ and DCN in TW Hya have abundances comparable to the abundances of their main isotopologues, HCO$^+$ and HCN [7], while cosmic abundance ratio of D/H is only 10$^{-5}$. It remains to be verified whether such a large degree of deuterium fractionation is a heritage of pre-disk cold evolutionary phase or these species are freshly produced in cold disk regions.

The most remarkable finding is that gas-phase molecular abundances in PPDs are lower by factors 5-100 compared to the values measured in the Taurus Molecular Cloud. This was interpreted as a joint action of intense UV and X-rays dissociation of molecules at elevated disk heights by the young central stars and due to proficient sticking of gas-phase molecules to dust grains in cold disk interiors.

## Inner disk: infrared spectroscopy

The results from the Infrared Space Observatory and the Spitzer Space Observatory have revealed the presence of a large amount of ices and amorphous and crystalline silicates and PAHs in disks [9-12]. The PAH features at ~3-12μm probe the intensity and shape of the incident radiation field and density distribution of the upper disk. Only ~10% of PPDs around young Sun-like stars show aliphatic (C-H) and aromatic (C=C) PAH bands, and often these bands appear only in outer disk zone (beyond ~30-100 AU). The small PAHs with unbound geometry (<50 C-atoms) are destroyed by the intense FUV and, in particular, X-ray stellar radiation, while bigger, ball-like compact PAHs do survive in PPDs. The PAHs remain in the gas-phase only in dilute, ice-free upper disk regions, and are frozen out in disk interiors. It has been speculated that these PAHs can be chemically transformed by irradiation/high temperature into a tar, kerogene-like material that is omni-present in carbonaceous chondritic meteorites. It is interesting to note that the 3μm diamond emission tends to be centrally-peaked in PPDs, which may indicate a high-temperature conversion of PAHs into nano-diamonds [25].

A multitude of amorphous and crystalline silicates with varying Fe/Mg ratios and grain topology/sizes have also been detected [26]. A striking result is that both the crystallinity fraction and the abundance ratio of forsterite/enstatite changes across a disk. The fact that mid-IR features are characteristic of mostly iron-poor silicates imply their rapid condensation from the gas-phase during transformation of a molecular cloud into a disk. Other metals like Fe and Ca are found as a component (often as sulfides and oxides) of glassy inclusions inside silicate matrix in meteorites. A variety of observed silicate band profiles is due to various average size of emitting grains, implying that small sub-micron-sized particles and bigger, several micron-sized grains exist in PPDs.

Ices are present in cold, outer disk region beyond the "snowline", at temperatures below ~120 K (r ≳ 20 AU). Their spectral features are easier to detect in highly-inclined systems. The major mantle component is water ice, with an abundance of about $10^{-4}$ (relative to the total particle density). Water ice is intermixed with other, more volatile ices of e.g. CO, $CO_2$, $NH_3$, $CH_4$, $H_2CO$, and HCOOH [27]. Typical abundances of these minor constituents are about 0.5-10% wrt to water. As it has been revealed from a complex shape of icy features having substructures, a substantial part of these volatile ices can be chemically bound to grain surface sites ("chemisorbed"), and thus will remain frozen even at elevated grain temperatures and UV irradiation.

The detection of NeII line emission from several disks, as well those from iron, sulfur, and argon, has confirmed theoretical predictions that the disk atmosphere is heavily ionized and superheated by the energetic photons. Detected rotational $H_2$ and (ro-)vibrational CO, $CO_2$, $C_2H_2$, HCN as well as $H_2O$ and OH lines trace a hot gas in inner, planet-forming disk zone with T ~300-1000 K [28]. The HCN to $C_2H_2$ ratio is a sensitive tracer of the UV/X-ray intensity or accretion rate inside an inner disk, while the easily populated CO lines allow to constrain kinetic temperature. The rich chemical composition inside planet-forming disk zones suggest that many endothermic reactions with barriers become active there.

## Solar nebula: cosmochemistry studies

The major result from analysis of meteoritic records is radioisotopic chronology and relative ages of various types of condensates. A substantial fraction of solids has been evaporated and re-condensed out during formation of the inner Solar nebula at thermodynamical equilibrium and thus follows a condensation sequence. This re-condensation of refractory matter has led to the chemically homogeneous composition of the matter of the inner Solar nebula (at bulk level), though some of presolar refractory grain materials have survived. The radial condensation gradient from the most refractory elements (e. g., metals ) to more volatile materials (e. g., ices) has been imprinted into composition of inner terrestrial rocky and outer gas-dominated Jovian planets and ice-rich comets in the Solar system. Meteorites have been formed in the inner nebular region, and as such have lost a substantial portion of volatile materials. Volatile elements are depleted in almost all materials from the planetary zone of the Solar System, and thus it is chemically different from the bulk primordial composition retained in the outer region, where comets formed. Thus the inner Solar Nebula does not retain memories of the pristine chemical composition [29].

The refractory droplets (chondrules) as well as metals and metal oxides (Calcium-Aluminum-rich Inclusions, CAIs, and metal alloys) have condensed first out of the gas at ~1500-1800 K and are among the oldest solids in the Solar system [~4.55 Gyr; see 30]. The CAIs have condensed prior to chondrules and likely survived many high-temperature rises, while chondrules have formed during a single melting event, followed by silicates (<1500 K), troilite (~700 K), and carbonaceous materials (~10-500 K). Thermodynamics of silicate condensation favors production of Mg-rich crystalline silicates at high temperatures followed by rapid gas cooling, as at lower temperatures iron becomes incorporated into the silicates. The chemical evolution of the outer part of the nebula, where ices have condensed, has been controlled by slow chemical kinetics and is harder to reconstruct without advanced theoretical modeling. The production of carbonaceous matter by chemical kinetics processes and disk dynamics is particularly sensitive to local conditions in the inner nebula, as combustion and pyrolysis of precursor hydrocarbons at high temperatures led to production of kerogene-like (mainly aromatic) carbonaceous materials in meteoritic samples [31]. Organic matter in meteorites, including amino acids and their precursor species, have likely formed from pristine ices upon steady temperature increase or UV/X-ray irradiation when icy grains have been transported inward by viscous disk evolution [32].

While chemical homogenization of the bulk matter in the inner Solar nebula has been caused by re-condensation processes, isotopic analysis of high-temperature condensates in chondritic meteorites shows that macroscopically significant oxygen isotopic anomalies still persist [33]. Anomalous isotopic fractionation can be caused either by mass-dependent (e. g., chemical reaction) or by mass-independent (e. g., photodissociation) processes. The oxygen isotopic anomalies have mainly been attributed to selective photochemistry of the CO isotopologues, where less abundant $^{13}CO$, $C^{18}O$, and $C^{17}O$ are easier to dissociate by the FUV photons, resulting in higher abundances of the $^{13}C$- and $^{17,18}O$-bearing species. Isotopic anomalies in the short-lived radio nuclides $^{26}Al$ and $^{41}Ca$ and other heavy elements in certain groups of meteorites point to minor temporal and spatial heterogeneity in the Solar nebula on a local scale. The source of this anomaly is either injection from a nearby supernova when Solar nebula was a few Myr old or due to irradiation by energetic cosmic ray particles [34].

The pristine materials from the planet-formation epoch of the Solar system have been preserved in comets and outer planets and their satellites. With *Galileo* and *Cassini*, the robotic missions to Jupiter and Saturn, better constrains on thermal structure and chemical diversity of atmospheres of these gas giants and their icy moons have been placed [35]. A complex hydrocarbon chemistry of the Titan's atmosphere and on it's surface has been found, which is active at extremely low temperatures of ~20 K. The comets have been assembled around or beyond Neptune and expelled gravitationally outward, and thus volatile C-, H-, O-containing ices contribute a large fraction of the cometary composition. They are considered as one of the most important sources of water delivered to young Earth. The recent *Stardust* space mission has returned the first samples of cometary dust grains, which show a complex structure of high-temperature solids (like crystalline silicates) embedded in low-temperature condensates, and which thus do not resemble pristine ISM dust [36]. The *Deep Impact* mission that crashed into the Tempel 1 comet allowed probing structural properties and composition of the ejected comet crust, with a higher dust-to-gas ratio than expected. A major puzzling fact is the presence of crystalline silicates in cometary dust, which require annealing of the amorphous silicates at temperatures above 800 K [37]. This can be explained by crystallization of silicates in the inner nebula, followed by outward radial transport [38] or due to shock-heating during the transition from a natal cloud to the nebula [39]. The low-temperature conditions at which comets have been assembled are imprinted into elevated D/H ratio, which has been observed in heavy water in the comet 8P/Tuttle. Observations of a few other comets have revealed the presence of some organic materials like HCN, $CH_4$ and $CH_3OH$. The simplest amino acid, glycine, has been recently discovered in samples of dust returned by *Stardust*, with isotopic composition of extraterrestrial origin.

## Theoretical picture of chemical structure of a disk

Table 2: Key chemical processes in protoplanetary disks

| Name | Representation | Example | Rate* |
|---|---|---|---|
| Radiative association | A + B → AB + v | $C^+ + H_2 → CH_2^+$ | $\sim 10^{-10}-10^{-17}$ cm$^3$ s$^{-1}$ |
| Ion-molecule | $A^+ + B → C^+ + D$ | $CO + H_3^+ → HCO^+ + H_2$ | $\sim 10^{-7}-10^{-10}$ cm$^3$ s$^{-1}$ |
| Neutral-neutral | A + B → C + D | $O + CH_3 → H_2CO + H$ | $\sim 10^{-10}-10^{-16}$ cm$^3$ s$^{-1}$ |
| Charge transfer | $A^+ + B → B^+ + C$ | $C^+ + Mg → C + Mg^+$ | $\sim 10^{-9}$ cm$^3$ s$^{-1}$ |
| Radiative recombination | $A^+ + e^- → A + v$ | $Mg^+ + e^- → Mg + v$ | $\sim 10^{-12}$ cm$^3$ s$^{-1}$ |
| Dissociative recombination | $AB^+ + e^- → A + B$ | $HCO^+ + e^- → CO + H$ | $\sim 10^{-7}$ cm$^3$ s$^{-1}$ |
| Ionization | $A + hv → A^+ + e^-$ | $C + hv → C^+ + e^-$ | $\sim 10^{-10}$ x RF** cm$^3$ s$^{-1}$ |
| Dissociation | AB + hv → A + B | CO + hv → C + O | $\sim 10^{-10}$ x RF cm$^3$ s$^{-1}$ |
| Accretion | A + g → A(g) | $H_2O + g → H_2O(g)$ | $\gtrsim 4\ 10^{-6}$ cm$^3$ s$^{-1}$ |
| Surface reaction | A(g) + B(g) → AB(g) | $H(g) + H(g) → H_2(g)$ | $\gtrsim 10^{-9}$ cm$^3$ s$^{-1}$ |
| Desorption | A(g) → A (hv or T) | $H_2CO(g) → H_2CO$ | $\sim 0-10^5$ s$^{-1}$ |

\* - chemical reaction rates vary widely in disks due to strong gradients of physical conditions,
\*\* - RF designates the local UV radiation field in the disk.

We present a general scheme of a PPD structure based on observational results and theoretical predictions in Fig. 2 [23, 40-44]. The viscous evolution of the disk gas is driven by the angular momentum transport due to turbulent torques. This leads to inward transport of the bulk of disk matter (~80%), while a minor fraction of the disk gas carries the angular momentum outward (mostly along midplane). This leads to a steady-state power-law surface density profile that is roughly $\Sigma(r) \propto r^{-3/2}$ (~1700 g cm$^{-2}$ at 1 AU for the Solar nebula model). In vertical direction, the gas density drops exponentially, as in the Earth atmosphere. The average temperature in disks decreases outward, and, for a given distance, increases from disk midplane to disk atmosphere. The stellar radiation heats up the upper layer such that the disk becomes flaring. The disk atmosphere has low density, resulting in higher gas temperature compared to dust temperature. Closer to the star gas temperature in the atmosphere rises so high (T ~ 1000-5000 K) that the matter starts to evaporate. Below the atmosphere disk density increases and the gas and dust temperatures reach an equilibrium value. In the very inner part, the disk midplane is again warmed up by the viscous accretion heating.

The intensity of ionizing and dissociating UV radiation is determined by the dust opacities, while penetration of the stellar X-ray radiation into the disk is controlled by the gas density structure. The larger the energy of a photon or a particle, the deeper disk regions it can reach. The X-ray emitting source is often located high above the stellar photosphere, at distances of several stellar radii (~0.01 AU), and thus X-ray photons reach the disk atmosphere at an oblique angle and able to penetrate deeper into the inner disk interior compared to the UV photons. The stellar UV intensity drops down as r$^{-2}$, while the X-ray intensity decreases even faster, and the interstellar UV radiation start to prevail in outer disk region (r $\gtrsim$ 100 AU). The FUV continuum radiation is blocked by a gas column of <0.01 g cm$^{-2}$, hard X-ray photons (~1-5 keV) are stopped at a gas column of 0.1-1 g cm$^{-2}$, and ~ GeV cosmic ray particles penetrate as large gas columns as ~100 g cm$^{-2}$.

Consequently, the disk structure can be divided into 4 chemically-distinct regions: the warm "inner zone" where a planetary system is built (observable currently only at IR and optical wavelengths), and outer disk region observed by modern radio-interferometric facilities. The inner region is at thermodynamical equilibrium, and chemical abundances there can be calculated at predefined temperature, pressure, and elemental abundances by condensation sequence modeling. The major problem here is to obtain reliable equilibrium coefficients for all species of interest. The chemical kinetics controls the evolution of the outer disk region (r > 20 – 50 AU). It is divided on a cold, dense disk midplane where dust grains are coated with ices, a slightly UV/X-ray-irradiated warm layer above the midplane with rich chemical composition, and a heavily ionized, hot and dilute disk atmosphere.

The key reactions for disk chemistry are summarized in Table 2. Modern astrochemical databases include up to 600 species involved in 6000 gas-phase, gas-grain, and surface reactions. Only ~10% of reaction rates have been accurately measured in the laboratory or calculated theoretically, making calculated molecular concentrations uncertain by factor of ~3-5 [45, 46]. The most widely utilized astrochemical networks: the University of Manchester (UMIST) [47], the Ohio State University (OSU) [48], and a network incorporated in the Meudon code [49]. Several datasets of surface reactions have also been compiled [50-52].

In the disk inner zone, three-body gas-phase processes become competitive at densities above $\sim 10^{10}$ cm$^{-3}$ [40]; otherwise all reactions in PPDs are two-body processes. First, light species (e. g., CH$^+$) and molecular bonds are formed by slow radiative association of atomic species, where an excess of energy required to stabilize a new product cannot be rapidly released as a photon. When polyatomic molecular radicals have formed, their radiative association chemistry proceeds at faster rate as these bigger species form a large unstable molecule in an excited state with a dense level structure, allowing emission of multiple photons [53]. The primal radiative association reaction that begins the molecular evolution at cosmic conditions is $C^+ + H_2 \rightarrow CH_2^+$. Rates of many slow radiative association processes are poorly determined.

Next, reactions between neutral and charged molecules begin. The ion-molecule reactions dominate chemistry, especially at low temperature in the outer disk region. Most of these reactions are exothermic, with high rate coefficients, $\sim 10^{-7}$ cm$^3$ s$^{-1}$ that may increase toward low temperatures, and low (if at all) reaction barriers [54]. The long-distance Coulomb attraction of an ion and a molecule with a high dipole moment is what makes the ion-molecule rates to increase when temperature decreases. Ion-neutral reactions lead to restructuring of molecular bonds. One of the most important reactions of this class in disks are protonation reactions, such as $H_3^+ + CO \rightarrow HCO^+ + H_2$. A number of neutral-neutral reactions between radicals and radicals, radicals and open-shell atoms, and radicals and unsaturated molecules are competitive in disks even at low temperatures, with a typical rate coefficient that is only about an order of magnitude lower than for the ion-molecule processes [48]. These reactions usually have barriers and as such become more active within the warm inner zone of a PPD. One of the most interesting reactions of this type is formation of formaldehyde: $CH_3 + O \rightarrow H_2CO + H$ [47]. The reaction rates of ion-molecule reactions are usually accurate within 50%, while those for neutral-neutral reactions are harder to constrain due to reaction barriers.

Rapid neutralization of polyatomic molecular ions by dissociative recombination with electrons and negative ions leads to production of smaller neutral molecules in disks. These processes are fast at low temperatures, with typical rates of about $10^{-7}$ cm$^3$ s$^{-1}$ [47]. For many observed molecules, dissociative recombination is an important formation pathway. Often, at late evolutionary phase, ~0.1 Myr, dissociative recombination balances protonation reactions, e.g. $H_3^+ + CO \rightarrow HCO^+ + H_2$ followed by $HCO^+ + e^- \rightarrow CO + H$. Dissociative recombination rates are not difficult to derive but branching ratios and products are not easily predictable.

The additional energy is brought to reacting gas species by cosmic ray particles, X-rays, and UV photons, which dissociate and ionize species, destroying molecular bonds. The CRPs dominate ionization at disk interiors (unless surface density does not exceed ~100 g cm$^{-2}$), whereas X-rays and FUV photons are important at upper disk regions (surface density ≲ 1 g cm$^{-2}$). The CRP- and X-ray-driven ionization is mainly caused by secondary energetic electrons upon primal ionization of H$_2$. The respective reaction rates are uncertain since the CRP and X-ray spectra cannot be directly derived from disk observations (e. g., due to local absorption of low-energy CRPs and X-rays). The key ionization reactions in PPDs are CRP/X-ray ionization of $H_2$ and He.

The ionization and dissociation of some chemical species proceed by absorption of the UV continuum radiation, while other molecules are destroyed by absorbing UV photons of particular energies. For example, $H_2$ and CO dissociate via absorptions by FUV photons in discrete lines, whereas other molecules are dissociated either by the continuum (e.g., $CH_4$) or by the continuum and lines (e.g., $C_2$; see 55, 56). Atmospheres of young Sun-like stars generate non-thermal UV radiation, with an intensity at 100 AU from the star that can be as high as 1000 (in units of the interstellar UV field, see [23]). In some stars like DM Tau a significant fraction of the total UV luminosity of the star is emitted in the Ly$\alpha$ line. Stellar and interstellar UV photons are able to penetrate deep into flaring disks by scattering on dust grains [42]. By dissociating and ionizing molecules, mild UV radiation drives active chemistry involving these ions and radicals, increasing chemical complexity of the disk. For example, the high ratio of CN to HCN abundances observed in disks is attributed to the intense photodissociation of HCN when part of the stellar UV flux comes as Ly$\alpha$ photons (121.6 nm; see [23]).

The key photodissociation reaction in disks is that for CO - a molecule that otherwise locks almost all elemental carbon from chemistry (see Table 2). Since the dissociation of the $H_2$ and CO molecules involves absorption of FUV at discrete wavelengths shortward of ~115 nm, self-shielding and isotope selective photodissociation effects are possible. In the disks, $H_2$ and CO are so abundant that dissociating FUV lines saturate, shielding the rest of molecules from being destroyed [57, 58]. In contrast, rare, less abundant isotopologues of CO absorb FUV photons at shifted wavelengths compared to $^{12}C^{16}O$, making self-shielding unimportant, and enriching the gas with rare O- and C-isotopes [59]. This is one of viable explanations of the oxygen isotopic anomalies found in various types of Solar nebula materials.

The observed overabundance of deuterated species in disks compared to the measured interstellar D/H ratio of ~$10^{-5}$ is well established. Deuterium enrichment operates at low temperatures of 10-20 K via a key mass-dependent fractionation reaction: $H_3^+ + HD \leftrightarrow H_2D^+ + H_2 + 232$ K [60]. A reactive $H_2D^+$ ion acts similar to $H_3^+$, donating D to neutral radicals via a "protonation" reaction. For example, predominant reaction pathway to produce $DCO^+$ is via ion-molecule reaction of CO with $H_2D^+$. Two other key fractionation reactions are active till higher temperatures (up to 70 K): $CH_3^+ + HD \leftrightarrow CH_2D + H_2 + 390$ K and $C_2H_2^+ + HD \leftrightarrow C_2HD^+ + H_2 + 550$ K [61]. The mass-dependent fractionation for heavier elements like C and O is not that effective due to much lower mass difference between isotopologues, and other mechanisms, like mass-independent isotope selective dissociation and surface processes have to be considered.

In cold disk regions, where T ~ 10-120 K, volatile species can condense out on dust grains. The rate of this freeze-out process depends on grain sizes and concentration, and is effective in outer disk midplane where grains do not grow significantly beyond ~1$\mu$m. At ~10-20 K many gas-phase molecules stick to a grain with a nearly 100% probability by weak van der Waals forces, while some may form a much stronger chemical bond with the surface. Consequently, desorption of chemisorbed species requires higher temperatures/UV irradiation than that of physisorbed molecules, and thus allow limited surface chemistry to occur even at harsh conditions.

Frozen species are released back to the gas by thermal evaporation, CRP-heating, and UV-desorption. Molecule evaporates thermally when it has energy to overcome binding to the surface [62]. Typical binding energies are about 1000 K for light molecules such as CO and $N_2$ [63], and larger for heavier cyanopolyynes and carbon chains (~5000 K). CRP-induced desorption is a transient heating event when a relativistic Fe hits a grain and rises T to 70 K, blowing up the ice mantle. Finally, the UV photons can kick off surface molecules with certain probability, which has been roughly measured to be ~$10^{-6}$-$10^{-3}$ [64].

The dust grain surfaces serve as a catalyst for many reactions with slow gas-phase rates. The most notable example is formation of molecular hydrogen that proceeds entirely on dust at T < 15-20 K [65]. A reaction may occur when an accreted atom or light radical, if it is not chemisorbed or desorbed back to the gas, hopes over the surface sites and finds a radical. At low temperatures surface hydrogenation of radicals is a key mechanism, leading to formation of saturated products like water, methane, ammonia, and methanol [50]. At higher temperatures of ~50 K heavier reactants become mobile (such as O, C, OH, HCO, etc.), and complex molecules are produced. So far this is the only viable route to form organic matter in protoplanetary disks, perceived by astrochemists.

# Future Directions of Research

The ALMA interferometer that is under construction in northern Chile will start early science observations in summer 2011 and become fully operational in 2012. It will open a new window and revolutionize our understanding of the Universe and protoplanetary disks in particular with its ultra high spatial resolution (up to $0.01''$), enormous collecting area (5000 m$^2$) and high sensitivity, and a broad frequency coverage (86-950 GHz or ~0.2-2mm). This is a gain by a factor of 50 in resolving power and orders of magnitude in sensitivity in comparison with current interferometers. First, this will allow us to observe high-lying rotational transitions of simple abundant species like HCO$^+$ and to discover more complex molecules with weaker lines at low frequencies. With this information we can better constrain disk chemical composition and physical conditions. Second, with much higher spatial resolution we will be able to image planet-forming regions around nearby PPDs in detail and thus determine its kinematics and distribution of dust emission. This is an essential step to determine if proto-planets are forming in disks, and to constrain gas dynamics through the disk (accretion flows, turbulence). Third, for inclined systems molecular layers will become directly visible for the first time, allowing us to better tune theoretical models and laboratory measurement regarding photodissociation and freeze-out [66]. Fourth, data acquisition will be fast with ALMA, and a statistically representative sample of protoplanetary disks around various young stars will be investigated. Other future instruments, like Square Kilometer Array (SKA) will allow detection of low-energy lines from highly complex (organic) species in cold disk regions at cm wavelengths, while James Webb Space Telescope (JWST) operating at IR will image disks in (ro-)vibrational molecular signatures, including biomarkers such as $CO_2$ and $CH_4$. A future progress in our understanding of the topic will not be achieved without development of more consistent, advanced chemical disk models that will include dynamical evolution of the disk, grain evolution, accurate transport of UV and X-rays, and more extended chemical networks. The latter will strongly benefit from on-going activities of laboratory/theoretical chemists to derive new rates and products of astrophysically-relevant processes, binding energies and photodesorption yields of surface species, and high-precision IR/mm spectra of potentially detectable, chemically interesting species.

# See also

[Add Cross-references to related entries in the reference work; the complete list of all entries could be found in OESYS (on line editorial system) by going to "download current List of Contributions" as a PDF document.

Please enter cross-references here]

→ …

→ …

# References and Further Reading

1. Lin, D. N. C. & Papaloizou, J. (1980), MNRAS, 191, 37
2. Payne, M. J. & Lodato, G. (2007), MNRAS, 381, 1597
3. http://astrochemistry.net
4. Dutrey, A., Guilloteau, S., & Guelin, M. (1997), A&A, 317, L55
5. Kastner, J. H., Zuckerman, B., Weintraub, D. A., & Forveille, T. (1997), Science, 277, 67
6. Aikawa, Y., Momose, M., Thi, W.-F., et al. (2003), PASJ, 55, 11
7. Qi, C., Wilner, D. J., Aikawa, Y., Blake, G. A., & Hogerheijde, M. R. (2008). ApJ, 681, 1396–1407
8. Bitner, M. A., Richter, M. J., Lacy, J. H., et al. (2007), ApJ, 661, L69

9. Pontoppidan, K. M., Dullemond, C. P., van Dishoeck, E. F., et al. (2005), ApJ, 622, 463
10. Lahuis, F., van Dishoeck, E. F., Boogert, A. C. A., et al. (2006), ApJ, 636, L145
11. Pascucci, I., Hollenbach, D., Najita, J., et al. (2007), ApJ, 663, 383
12. van Dishoeck, E. F. (2004), ARA&A, 42, 119
13. Grossman, L. & Larimer, J. W. (1974). Reviews of Geophysics and Space Physics, 12, 71–101
14. Mumma, M. J., Weissman, P. R., & Stern, S. A. (1993), in Protostars and Planets III, ed. E. H. Levy & J. I. Lunine, 1177–1252
15. Lauretta, D. S. & McSween, Jr., H. Y. (2006), Meteorites and the Early Solar System II (Meteorites and the Early Solar System II)
16. McSween, Jr., H. Y. & Huss, G. R. (2010), Cosmochemistry (Cambridge Univ. Press)
17. Rodmann, J., Henning, T., Chandler, C. J., Mundy, L. G., & Wilner, D. J. (2006). A&A, 446, 211–221
18. Cortes, S. R., Meyer, M. R., Carpenter, J. M., et al. (2009). ApJ, 697, 1305–1315
19. Dartois, E., Dutrey, A., & Guilloteau, S. (2003). A&A, 399, 773–787
20. Piétu, V., Dutrey, A., & Guilloteau, S. (2007). A&A, 467, 163–178
21. Qi, C., & et al. (2006), ApJL, 636, L157
22. Dutrey, A., Henning, T., Guilloteau, S., et al. (2007). A&A, 464, 615–623
23. Bergin, E., Calvet, N., D'Alessio, P., & Herczeg, G. J. (2003). ApJ, 591, L159–L162
24. Henning, T., et al. (2010), ApJ, 714, 1511
25. Goto, M., et al. (2009), ApJ, 693, 610
26. Bouwman, J., Henning, T., Hillenbrand, L. A., et al. (2008). ApJ, 683, 479–498
27. Zasowski, G., Kemper, F., Watson, D. M., et al. (2009). ApJ, 694, 459–478
28. Salyk, C., Pontoppidan, K. M., Blake, G. A., et al. (2008). ApJ, 676, L49–L52
29. Trieloff, M. & Palme, H. (2006), The origin of solids in the early Solar System (Planet Formation), 64
30. Day, J. M. D., Ash, R. D., Liu, Y., et al. (2009). Nature, 457, 179–182
31. Morgan, Jr., W. A., Feigelson, E. D., Wang, H., & Frenklach, M. (1991). Science, 252, 109–112
32. Ehrenfreund, P. & Charnley, S. B. (2000). ARA&A, 38, 427–483
33. Clayton, R. N. (2007). Annual Review of Earth and Planetary Sciences, 35, 1–19
34. Goswami, J. N. & Vanhala, H. A. T. (2000), in Prostostars and Planets IV, ed. V. Mannings, A. P. Boss, & R. S. S., 963–990
35. Niemann, H. B., Atreya, S. K., Bauer, S. J., et al. (2005). Nature, 438, 779–784
36. Brownlee, D. E., Horz, F., Newburn, R. L., et al. (2004). Science, 304, 1764–1769
37. Wooden, D., Desch, S., Harker, D., Gail, H.-P., & Keller, L. (2007), in Protostars and Planets V, ed. B. Reipurth, D. Jewitt, & K. Keil, 815–833
38. Gail, H.-P. (2001). A&A, 378, 192–213
39. Ábrahám, P., Juhász, A., Dullemond, C. P., et al. (2009). Nature, 459, 224–226
40. Aikawa, Y. & Herbst, E. (1999), A&A, 351, 233
41. Willacy, K. & Langer, W. D. (2000), ApJ, 544, 903
42. van Zadelhoff, G.-J., Aikawa, Y., Hogerheijde, M. R., & van Dishoeck, E. F. (2003), A&A, 397, 789
43. Semenov, D., Wiebe, D., & Henning, T. (2004), A&A, 417, 93
44. Bergin, E. A., Aikawa, Y., Blake, G. A., & van Dishoeck, E. F. (2007), in Protostars and Planets V, ed. B. Reipurth, D. Jewitt, & K. Keil, 751–766
45. Vasyunin, A. I., Semenov, D., Henning, T., et al. (2008), ApJ, 672, 629
46. Wakelam, V., Herbst, E., & Selsis, F. (2006), A&A, 451, 551
47. Woodall, J., Agúndez, M., Markwick-Kemper, A. J., & Millar, T. J. (2007), A&A, 466, 1197
48. Smith, I. W. M., Herbst, E., & Chang, Q. (2004), MNRAS, 350, 323
49. Le Petit, F., Nehmé, C., Le Bourlot, J., & Roueff, E. (2006), ApJS, 164, 506
50. Tielens, A. G. G. M. & Hagen, W. (1982), A&A, 114, 245
51. Hasegawa, T. I., Herbst, E., & Leung, C. M. (1992), ApJS, 82, 167
52. Garrod, R. T. & Herbst, E. (2006), A&A, 457, 927
53. Herbst, E. & Klemperer, W. (1973), ApJ, 185, 505
54. Dalgarno, A. & Black, J. H. (1976), Reports of Progress in Physics, 39, 573
55. van Dishoeck, E. F. (1988), in ASSL Vol. 146: Rate Coefficients in Astrochemistry, ed. T. Millar & D. Williams (Kluwer Academic Publishers, Dordrecht), 49–72

56. van Dishoeck, E. F., Jonkheid, B., & van Hemert, M. C. (2006), in Faraday discussion, Vol. 133, Chemical evolution of the Universe, ed. I. R. Sims & D. A. Williams, 231–244
57. Draine, B. T. & Bertoldi, F. (1996), ApJ, 468, 269
58. Lee, H.-H., Herbst, E., Pineau des Forets, G., Roueff, E., & Le Bourlot, J. (1996), A&A, 311, 690
59. Thiemens, M. H. & Heidenreich, III, J. E. (1983), Science, 219, 1073
60. Gerlich, D., Herbst, E., & Roueff, E. (2002), Planet. Space Sci., 50, 1275
61. Asvany, O., Schlemmer, S., & Gerlich, D. (2004), ApJ, 617, 685
62. Leger, A., Jura, M., & Omont, A. (1985), A&A, 144, 147
63. Bisschop, S. E., Fraser, H. J., Öberg, K. I., van Dishoeck, E. F., & Schlemmer, S. (2006), A&A, 449, 1297
64. Öberg, K. I., Fuchs, G. W., Awad, Z., et al. (2007), ApJ, 662, L23
65. Hollenbach, D. & Salpeter, E. E. (1971), ApJ, 163, 155
66. Semenov, D., Pavlyuchenkov, Y., Henning, T., Wolf, S., & Launhardt, R. (2008). ApJ, 673, L195–L198